\documentclass[onecollarge,natbib]{svjour2}
\bibpunct{[}{]}{;}{n}{}{,} 
\smartqed  
\usepackage{graphicx}

\usepackage{amssymb}
\newcommand{\beq}{\begin{eqnarray}}
\newcommand{\nneeq}{\nonumber \end{eqnarray}}
\newcommand{\eeq}{\end{eqnarray}}

\newcommand{\es}{& = &}
\newcommand{\rs}{\, = \,}

\newcommand{\cM}{ {\cal M} }
\newcommand{\cH}{ {\cal H} }

\newcommand{\cL}{ {\cal L} }

\journalname{Few-Body Systems}

\begin{document}

\title{ Hypothesis of quark binding by condensation of gluons in hadrons 
\footnote{Presented by the author at LIGHTCONE
2011, 23 - 27 May, 2011, Dallas.} }
\author{ Stanis{\l}aw D. G{\l}azek } 
\institute{S. D. G{\l}azek   \at
              Institute of Theoretical Physics, Faculty of Physics, University of Warsaw \\
              Tel.: +48 22 553 2000 \\
              Fax:  +48 22 621 9475 \\
              \email{stglazek@fuw.edu.pl} }
\date{Received: date / Accepted: date}

\maketitle

\begin{abstract}
Hypothesis of quark binding through condensation of gluons inside hadrons
is formulated in the context of a renormalization group procedure for 
effective particles (RGPEP) in the light-front (LF) Hamiltonian approach 
to QCD. At the momentum scales of relative motion of hadronic constituents that 
are comparable with $\Lambda_{QCD}$, the hypothetical boost-invariant constituent 
dynamics is identified using gauge symmetry. The resulting picture of mesons 
and baryons closely resembles constituent quark models with harmonic oscillator 
potentials, shares some features of AdS/QCD, and can be systematically 
studied using RGPEP in QCD.
\end{abstract}

\section{ Canonical approach to QCD }

The canonical plan for solving QCD is to start
from the corresponding classical gauge-invariant
Lagrangian density, $\cL_{can}$, derive from it a
candidate for a Hamiltonian density, $\cH_{can}$,
and evaluate the Hamiltonian $H_{can} = \int d^3x
\, \cH_{can} $, in which the quark and gluon
fields are quantized. Our discussion of the
quantum theory begins with the standard form of
Hamiltonian dynamics, called the instant form
(IF), but our goal is to address the front form
(FF) of the theory \cite{Dirac}. 

The canonical operator $H_{can}$ acts in the Fock 
space of states $|\psi\rangle$ that are built from 
a vacuum by the operators that create quarks and 
gluons. Then, a solution to the eigenvalue problem
\beq
H_{can}  |\psi \rangle \es E \, |\psi \rangle \, , 
\eeq
with energy of the form $E = \sqrt{ M^2_h + \vec P^{\,2} } $, 
where $M_h$ is a mass, should represent a hadron $h$
moving with momentum $\vec P$. Time evolution of all 
states would be described using operators of the form 
$U(t_2, t_1) = e^{-i H_{can} \, (t_2-t_1)}$, etc. The 
ultimate goal of such plan for QCD is to achieve a 
quality of the wave function representation of hadrons 
that matches, and in the future hopefully even exceeds 
the quality of today's QED representation of atoms
and their chemistry.

\section{ Time-honored problems of canonical approach }

The canonical approach encounters divergence
problems that are known for a long time to require
a logical resolution \cite{DiracDeadWood}. A hint
of the resolution was discovered also a long time
ago \cite{Wilson1965} through the concept of a
renormalization group (RG) procedure
\cite{Wilson1970}. The idea is that one can {\it
calculate} the counterterms required by the
adopted regularization. This is done in a sequence
of steps that eventually produce a manageable
effective theory. To begin with, one has to
replace $H_{can}$ \cite{LepageBrodsky,
SrivastavaBrodsky} with a regulated operator,
$H_{can}^\Delta$, where $\Delta$ stands for an
extreme cutoff parameter. The counterterms,
$CT^\Delta$, that need to be included in the
initial Hamiltonian, $H = H_{can}^\Delta +
CT^\Delta$, are obtained from the condition that
the matrix elements of the desired effective
Hamiltonian do not depend on the regularization in
$H$. The calculation of counterterms is based on
evaluation of a whole family of effective
Hamiltonians, $H_\lambda$ labeled by the RG
parameter $\lambda$, which can be chosen to have
the dimension of momentum. This parameter plays
the role of a sliding momentum cutoff. 

The question that so far has no satisfactory resolution 
in canonical approaches to QCD is: What precisely is the 
operator $H_\lambda$ in which $\lambda$ is so small that 
the eigenvalue problem for $H_\lambda$ can be solved 
using computers? One well-known reason for the difficulty 
is asymptotic freedom \cite{af1,af2}. It implies that the 
effective coupling constant $g_\lambda$ in $H_\lambda$ 
increases according to the rule $g_\lambda \sim 1/\ln 
(\lambda/\Lambda_{QCD})$ when $\lambda$ decreases (for 
the case of $H_\lambda$ in LF QCD, see \cite{gluons}), 
and becomes too large too soon for a perturbative procedure 
to produce $H_\lambda$ with a sufficiently small $\lambda$ 
for a reliable computation of quark and gluon wave functions 
of hadrons.

The chief difficulty of the IF of dynamics, however, is 
that we do not know the ground state of QCD. The canonical 
IF interaction Hamiltonians are able to create virtual 
particles from empty space and as time flows they generate 
this way an infinitely complex state that so far nobody 
can describe \cite{DiracDeadWood}. The ground state, called 
vacuum, should be invariant with respect to the Lorentz 
transformations. But such transformations change momenta 
by arbitrary amounts and a theory with a finite cutoff on 
momentum cannot have a Lorentz invariant ground state. 
Nevertheless, a covariant perturbation theory in QCD 
allows for introduction of non-trivial parameters
that can be interpretated as vacuum expectation values 
of operators. These parameters are related to the
spectrum of hadrons through dispersion relations. This 
is how the quark and gluon vacuum condensates are 
introduced in the QCD sum rules \cite{SVZ1,SVZ2}.

In contrast to the case of the IF of dynamics, it is 
well-known that the FF of Hamiltonian dynamics necessarily 
leads to a trivial vacuum state, apparently allowing one
to make progress without getting stuck in the hard vacuum
problem right away. This fact motivated efforts to set up 
a similarity renormalization group (SRG) procedure for 
Hamiltonians \cite{SRG1, SRG2} and apply it to LF QCD 
\cite{Wilson1994}. The effects that in the IF are associated 
with the unknown state of the vacuum, were suggested in 
Ref. \cite{Wilson1994} to be contained in the new terms in 
$H$, and thus also in $H_\lambda$, that a precise SRG procedure 
could identify. Through later efforts, summarized in Ref. 
\cite{Glazek2011}, it was found that there exists an 
additional possibility for the vacuum-like terms to emerge 
in LF QCD in a systematic calculation. The additional 
possibility is the subject of this article.

\section{ Renormalization group procedure for effective particles }

In the renormalization group procedure for effective 
particles (RGPEP) that is used in Ref. \cite{Glazek2011}
to suggest a new locus for vacuum-like condensate
effects in LF QCD, the key element of reasoning is
a unitary connection between creation and annihilation 
operators for bare, canonical quarks and gluons,
and the operators for effective particles,
\beq
\label{Us}
a_s \rs U_s \, a_0 \, U_s^\dagger \, .
\eeq
The parameter $s = 1/\lambda$ has interpretation
of the size of the effective particles with respect 
to strong interactions. The parameter $\lambda$ 
describes the momentum width of vertex form factors
in interaction terms, and $s$, as its inverse, 
corresponds to the non-local interaction vertex 
range in space. Thus, the subscript 0 refers to 
the point-like particles, with only local interactions,
while $s>0$ implies non-local interactions \cite{Glazek2010}.
Correspondingly, the effective quantum fields in space
are constructed according to the well-known Fourier
superposition rule \cite{Glazek2011}
\beq
\psi_s(x)  \rs  \int [p] \, a_{s p}     \, e^{ - i
p \, x} \, .
\eeq 
As a result, $\psi_0(x)$ corresponds to canonical 
fields, and $\psi_s(x)$ to fields of effective
quanta of size $s$. By writing $\psi_s(x) = \psi(x,s)$,
one can realize that the scale-dependent effective
theories form together a single 5-dimensional theory, 
the 5th dimension being the size of effective particles. 
It is natural to expect that this size is dynamically 
limited from above by $s_{QCD} \sim 1/\Lambda_{QCD}$ and 
that $s_{QCD}$ corresponds to the depth of bulk penetration 
by matter fields in AdS/QCD models. We shall comment on
this issue near the end of the article. 

The transformation $U_s$ in Eq. (\ref{Us}) is
constructed in such a way (see \cite{Glazek2011}
and references therein for computational details) 
that the Hamiltonian is not changed, $H_s(a_s) = 
H(a_0)$. But the Hamiltonians differ in their 
structure: $H_s(a_s)$ is a combination of products 
of operators $a_s$ with coefficients $c_s$ that are 
different from the coefficients $c_0$ of corresponding 
products of operators $a_0$ in $H(a_0)$. RGPEP provides
differential (or algebraic) equations that produce
expressions for the coefficients $c_s$. From the
equality 
\beq 
H_s( a_0 ) 
\es 
U_s^\dagger H(a_0) \, U_s \, ,
\eeq
and the condition $U_0 = 1$, one obtains
\beq
\label{RGPEPH}
{d \over d s^4 } \, H_s( a_0 ) 
\es
[G_s, H_s(a_0)] \, , 
\eeq
with the generator 
\beq
G_s \es 
- U_s^\dagger {d \over d s^4 } \,
U_s 
\eeq
and initial condition $H_0(a_0) = H(a_0) = H_{can}^\Delta 
+ CT^\Delta$. The use of $s^4$ is due to dimensional 
reasons since $G_s$ is designed to have dimension of 
mass to power 4. The non-perturbative, boost invariant 
generator $G_s$ for RGPEP is given in Eq. (C.1) of 
Ref. \cite{Glazek2011}. It is expressed in terms of 
the operators $a_0$ in the form of a commutator,
\beq
\label{Generator}
G_s \es [H_{free}, H^+_s] \, , 
\eeq
where $H_{free}$ denotes the part of $H(a_0)$ that 
involves only products of the form $a_0^\dagger
a_0$ (one particle operators). $H^+_s$ is the 
remaining part of the Hamiltonian. The superscript
$+$ indicates that the coefficients $c_s$ in each and 
every term in $H^+_s$ are multiplied by the square 
of the total $+$ momentum of the particles that
participate in the interaction described by a given 
term (see Eqs. (C.2) and (C.3) in \cite{Glazek2011}). 
As a result, when $s$ increases, one obtains from Eq. 
(\ref{RGPEPH}) the effective interactions that are 
increasingly tempered by the vertex form factors that 
limit the changes of invariant masses of interacting 
effective particles. For example, solving Eq. 
(\ref{RGPEPH}) in a lowest order in powers of the 
interaction strength for any interaction term $V$, 
one obtains
\beq
\langle m(s)|V_s(a_s)|n(s)\rangle 
\es
e^{ - s^4(\cM^2_{Im} - \cM^2_{In})^2} 
\langle m(s)|V_0(a_s)|n(s)\rangle \, , 
\eeq
where $|m(s)\rangle$ and $|n(s)\rangle$ denote 
arbitrary states in the Fock space built using 
operators $a_s$ and $\cM_{Im}$ and $\cM_{In}$ 
denote the total invariant masses of only these
subsets of effective particles in the corresponding 
states that are directly involved in the matrix 
element of the interaction $V$. The RGPEP vertex 
form factors suggest that the effective LF Fock 
space description of hadronic states may actually 
converge. This expectation needs to be verified 
by explicit calculations, which is a serious 
challenge.

Apparently similar to Eq. (\ref{RGPEPH}), beautiful 
flow equations have been developed by Wegner
\cite{Wegner,Kehrein}. The five main ways RGPEP 
differs are: (1) The transformation of a Hamiltonian 
in RGPEP is limited to transformations of creation
and annihilation operators, which is a narrower
class of transformations than rotating matrices, 
since matrices that result from evaluating matrix
elements of linear combinations of products of
rotated creation and annihilation operators form
only a subset in the set of all matrices of interest 
in quantum mechanics; (2) Eq. (\ref{RGPEPH}) does 
not require specification of the full diagonal 
matrix elements of the evolving Hamiltonian, as 
Wegner's equation does; (3) The Hamiltonian as an 
operator is not altered at all by Eq. (\ref{RGPEPH}) 
since the coefficients $c_s$ evolve in a way that 
is compensated by the evolution of operators $a_s$,
but the RGPEP derivation of counterterms alters 
the initial condition at $s=0$, a feature that
Wegner's equation does not include, since it
treats the initial Hamiltonian as given (the
calculation of $H$ is one of two generic goals of
RGPEP, as it is in the SRG \cite{SRG1,SRG2}, the 
other goal being the evaluation of corresponding 
effective Hamiltonians with $s$ sufficiently large 
so that they can be used in numerical computations,
and the latter goal is shared with Wegner's flows); 
(4) The RGPEP operator calculus renders coefficients 
$c_s$ in the Hamiltonians $H_s(a_s)$ that can be 
applied to arbitrary states in the Fock space, 
instead of only to a specified set of states that 
is used in defining the Wegner matrix equations; 
(5) The generator $G_s$ in Eq. (\ref{Generator}) 
is constructed to preserve the 7 kinematical 
symmetries of the FF of Hamiltonian dynamics. The
feature (5) is essential for application of RGPEP
to QCD since it allows one to kinematically connect 
the picture of a hadron in its rest frame with the 
picture of the same hadron in the infinite momentum 
frame. This is a prerequisite for any formulation 
of QCD that aims at simultaneously explaining the 
constituent quark model classification of hadrons 
in the particle data tables and the parton 
distribution functions measured in deep inelastic 
scattering processes as well as other high-energy 
hadronic properties that manifest themselves in 
collisions of fast moving hadrons.

\section{ Scale-dependent constituent picture and the gluon condensate in hadrons }

\begin{figure*}
\centering
\includegraphics[width=0.9\textwidth]{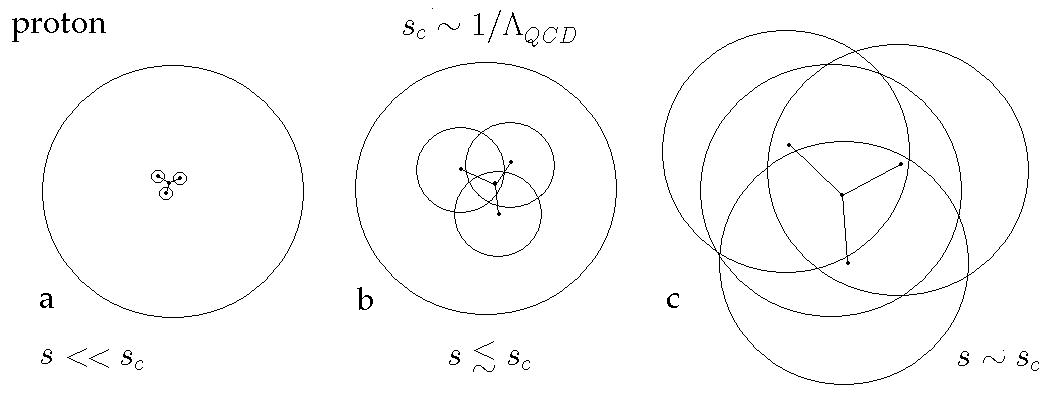}
\caption{
Visualization of the idea that a three-quark 
configuration in a proton evolves with the RGPEP 
scale parameter $s$: (a) the picture at $s$ much 
smaller than the scale $s_c$ that corresponds to 
the constituent model, (b) $s$ somewhat smaller 
than $s_c$, and (c) $s$ comparable with $s_c$. The 
key consequence of the RGPEP is that the slow 
effective quarks must be large. For $s \sim 1/
\Lambda_{QCD}$, they are as large as the proton 
itself. In the case (c), they overlap a lot and 
balance color to zero inside a large part of the 
proton volume.}
\label{fig:1}     
\end{figure*}

Operators $W_{s_2 s_1} = U_{s_2} U^\dagger_{s_1}$
transform a hadron state $| \psi \rangle$ constructed 
in terms of quarks and gluons of size $s_1$ into
the same state but constructed in terms of quarks and 
gluons of size $s_2$. Since the operators $U_s$ depend
on interactions, they change the number of virtual 
particles in states they act on. Consequently, $W_{s_2 
s_1}$ also changes the number of particles. Thus, even 
if one assumes that for $s_1 \sim s_c = 1/\Lambda_{QCD}$ 
a proton can be represented as built from just three
large constituent quarks, Fig. \ref{fig:1}c, a change of 
scale from $s \sim s_c$ to $s < s_c$ is necessarily 
associated with creation of additional virtual particles
whose presence is indicated in Figs. \ref{fig:1}b and  
\ref{fig:1}a by the large circle. Therefore, while
the eigenvalue problem of $H_{s_c}(a_{s_c})$ for the 
proton state may take the form of a Schr\"odinger equation 
for three constituent quarks in a potential well, 
the eigenvalue problem of $H_s(a_s)$ with $s < s_c$ for 
the same proton state must take the form of a Schr\"odinger 
equation for three smaller quarks that are accompanied by 
additional virtual gluons (we ignore in this discussion 
additional quark-anti-quark pairs). These gluons are 
condensed only in the volume of the proton. The expectation
value of the gluon field-strength operator squared in 
this cloud of gluons is described below as the parameter
that plays the role of a gluon condensate parameter in 
the eigenvalue problems of effective LF Hamiltonians 
derived in QCD using RGPEP. 

The title hypothesis of this article states that, 
although the FF vacuum is trivial, the hadronic gluon 
content is not, and it is the latter that provides 
the FF dynamical effects which are otherwise associated 
with the vacuum in the IF of dynamics. The remaining 
part of this article briefly describes results that 
support the title hypothesis.

\section{ Gauge symmetry and the effective dynamics for quarks of size $s \lesssim s_c$ }

The eigenvalue problems for mesons (M) and baryons (B) 
expressed in terms of quarks and gluons from Fig. 
\ref{fig:1}b, based on the relations $ W_{s s_c} |12 
\rangle_{s_c} = |12G \rangle_s  $ and $ W_{s s_c} 
|123\rangle_{s_c} = |123G\rangle_s  $, determine 
wave functions $\psi_s (12 G)$ and $\psi_s (123G)$
in the corresponding states, 
\beq
|M \rangle_{s  }  \rs  \sum_{12 G}  \psi_s (12 G)  \,|12G \rangle_{s } 
\quad , \quad \quad &&
|B \rangle_{s  }  \rs  \sum_{123G}  \psi_s (123G)
\,|123G\rangle_{s } \, ,
\eeq
where the component $G$ represents the gluons condensed 
in a hadron. The operator $G^\dagger$ that creates the
gluon component of size of a hadron from the LF vacuum 
is approximated by a creation operator for a scalar 
particle. The effective eigenvalue problems for mesons
and baryons, $H_s|h \rangle_s  \rs  E \, |h \rangle_s $,
are projected on the basis states $|12G \rangle_s$ and
$|123G \rangle_s$, respectively, in order to obtain the 
wave functions  $\psi_s (12 G)$ and $\psi_s (123G)$.
The basis states are constructed to include the 
color-transport factors between $i$-th quark and a geometrical 
center, $\underline{x}$, of quarks in a hadron, $T_i = 
\exp{-ig \int_{\underline{x}}^{x_i} dx_\mu A^\mu}$, and
the gluon field operator $A$ is approximated using the
Schwinger gauge by $A^\mu(x) = {1 \over 2} \, (x-x_G)_\nu 
G^{\nu \mu}$, where $x_G$ denotes the position of the 
gluon body created by $G^\dagger$ in the hadron and
$G^{\mu \nu}$ is the gluon field strength operator 
at this point (see \cite{Glazek2011,GlazekSchaden}).

Currently, in the absence of precise numerical
information about the true low-energy effective
Hamiltonian for LF QCD, local gauge symmetry is
used to establish the structure of $H_s(a_s)$ for
$s \lesssim s_c$ as follows \cite{Glazek2011}.
Center-of-mass motion of an eigenstate of the LF
Hamiltonian separates out completely from the
eigenvalue problem and the equation one is left
with is for an operator whose eigenvalue is the
mass squared of a hadron. The operator itself is a
sum of a free mass squared of constituents,
denoted here by $\cM^2$, plus the interactions
which we do not know, say $V$. But we know that in
the rest frame of constituent quarks their
invariant mass squared is the square of the sum of
their energies. These energies can be approximated
by making a non-relativistic (NR) expansion, $E_p
= m + p^2/(2m)$, since the RGPEP vertex form
factors for $s \lesssim s_c$ prevent the
interactions from accelerating constituent quarks
to large relative speeds. Thus, one can sum the
individual quark energies and square the sum,
neglecting terms smaller than $p^2$. The result is
a simple quadratic expression in relative momenta
of quarks. This result is easy to obtain for
arbitrary masses of individual quarks, so that one
knows the right coefficients with which the NR
momenta squared enter the invariant mass squared.
By comparison of this result with the exact
expression for the invariant mass squared in the
FF of dynamics, which is quadratic in the
transverse relative momenta but a more complicated
function of longitudinal momentum fractions
carried by quarks, one learns how to write the
latter in terms of the three-momenta known in the
NR theory using LF variables \cite{Glazek2011}.
Now the crux is that we also know how gauge
symmetry dictates interaction through the minimal
coupling in the NR theory: $\vec p \rightarrow
\vec p - g \vec A$. So, by analogy, we also know
how to estimate the effects of the minimal
coupling in the LF mass squared. Namely, we trace
the consequences of the minimal coupling in the NR
theory and introduce the corresponding effects in
the LF theory. As a result, we obtain a free
invariant mass squared of quarks plus interaction
terms induced by expectation values of the gluon
field in the gluon component $G$. Our result is
the gauge symmetry candidate for effective
$H_s(a_s)$. 

In short, the NR reasoning proceeds along the lines of 
Ref. \cite{GlazekSchaden}, except that on the basis of 
RGPEP on the LF one considers expectation values of 
$\vec A^2$ in the gluonic component of a hadron, $|G\rangle = 
G^\dagger |0\rangle$, rather than in the omnipresent 
vacuum $|\Omega \rangle$. After inclusion of the color
transport factors, in a crude Abelian mean-field 
approximation, the expectation values of the type 
$\langle G|g_s^2 \vec A^2 |G\rangle$, with $\vec A = \vec B 
\times \vec r /2$ and $\vec B$ being the magnetic part 
of $G^{\mu \nu}$, render harmonic potentials of the 
form $\varphi_G^2 \vec r^2$ where $\varphi^2_G =
\langle G|(\alpha_s/\pi) G^{\mu \nu c} G_{\mu \nu}^c 
|G \rangle/\langle G | G\rangle$ and $\vec r$ denotes 
the relative position of the quarks (only their relative 
position appears because of gauge symmetry). 

By definition, in the momentum representation, the relative 
distances are defined as gradients with respect to the 
relative momenta. But we already know how to identify these 
momenta through the NR approximation that holds in the smallest
mass eigenstates of RGPEP Hamiltonians with large $s$. 
Namely, in the case of mesons built from a quark of momentum 
$p_1$, anti-quark of momentum $p_2$ (both having the same 
constituent quark mass $m$) and the glue component $G$, 
using notation $P = p_1 + p_2$, $x = p_1^+/P^+$, $p_1^\perp 
= x P^\perp + \kappa^\perp$, $p_2^\perp = (1-x) P^\perp - 
\kappa^\perp$, one has \cite{Glazek2011}
\beq
\label{kperp}
k^\perp
\es 
{ \kappa^\perp \over 2\sqrt{ x(1-x)} } 
\quad , \quad \quad 
k^z
\rs 
{ 2x-1 \over 2\sqrt{ x(1-x)} } \, m \, ,
\eeq
and the effective quark dynamics in the gluon condensate 
inside a meson takes the form
\beq
\label{meson}
\cM^2_{q \bar q} 
\es 
4m^2 
+
4 \left[ \vec k^{\,2} + 
{1 \over 2} m^2 \left( {\pi \varphi_G \over 3m}
\right)^2
{1 \over 2} \left( i \, {\partial \over \partial
\vec k} \right)^2
\right] \, .
\eeq
For baryons one obtains (see
\cite{Glazek2011} for definitions of momenta 
$\vec K$ and $\vec Q$)
\beq
\label{baryon}
{\cal M}_{3q}^2 \es 9m^2 + 6 \, \vec K^{\,2} 
                         + {9 \over 2} \, \vec Q^{\,2}  
+ 3m^2 \left( {\pi \varphi_G \over 3 m }
\right)^2 {5 \over 8} \left[ {1 \over 2}  \left( i \, {\partial \over \partial
\vec K} \right)^2 + {2 \over 3 } \left( i \, {\partial \over \partial
\vec Q} \right)^2
\right] \, .
\eeq
The corresponding harmonic oscillator frequencies 
$\omega_M = \pi \varphi_G /(3m)$ and  
$\omega_B = \sqrt{ 5/8 } \,\, \omega_M$
match phenomenologically accurate constituent quark 
models (e.g., see \cite{CQM1,CQM2,CQM3})
provided that one numerically identifies $\varphi_G^2$ 
with the vacuum gluon condensate parameter $\langle 
\Omega| (\alpha_s/\pi) G^{\mu \nu c} G_{\mu \nu}^c |\Omega
\rangle/\langle \Omega|\Omega \rangle $ fitted to the 
spectrum of hadron masses in the QCD sum rules 
\cite{SVZ1,SVZ2}. We reinterpret the vacuum condensate 
parameter as corresponding to the gluons condensing 
only inside hadrons. The ground-state eigenfunctions 
of operators in Eqs. (\ref{meson}) and (\ref{baryon})
are exponentials of the invariant mass squared of
the effective quarks,
\beq
\label{finalGaussian}
\psi_n
\es
N \, 
\exp{ \left\{ - {1 \over 2 n m
\omega} \left[ \left( \sum_{i=1}^{n} p_i \right)^2
- (n m)^2 \right] \right\} } \, ,
\eeq
where $n = 2$ for mesons and $n = 3$ for baryons,
with $\omega = \omega_M$ and $\omega = \omega_B$,
respectively.

\section{ Observables }

Since the quarks are accompanied by the gluons,
i.e., they move with respect to the gluon body
$G$, the electroweak hadron form factors are
obtained in the form of the Fermi motion-smeared
form factors for the effective quarks alone.
However, the smearing is a small effect, so that
the resulting form factors closely resemble
results obtained in quark models. At the same
time, RGPEP appears to provide the right tools for
understanding a transition between the asymptotic
counting-rules for hard scattering processes and
the soft, non-perturbative effects in low-momentum
transfer processes. Similar comments apply in the
case of structure functions which depend on the
Bjorken $x$ and momentum transfer $Q$. In the
RGPEP, the transformation $W_{s_Q s_c}$ provides a
connection between the current quarks of a small
size $s_Q$, i.e., the ones that are capable of a
sudden absorption or emission of a hard-photon,
and the effective constituent quarks of size $s_c
\sim 1/\Lambda_{QCD}$. This transformation is
expected to describe the evolution of parton
distributions, universally in $Q$ and $x$ since
the size parameter controls dependence on the
invariant masses that are functions of $Q$ and $x$
simultaneously. Therefore, one may expect the same
RGPEP tools to help shed some light on the gluon
saturation mechanism in QCD as seen in the
inelastic, inclusive or semi-inclusive processes.

It is also worth observing that the momentum
variable $ k^\perp = \kappa^\perp/\sqrt{ x(1-x)} $
that Brodsky and de Teramond discovered in their
LF holographic AdS/QCD picture of hadrons
\cite{FAdS1,FAdS2,FAdS3}, appears to match the one
dictated here by gauge symmetry and the
condensation of gluons inside hadrons. This
matching warrants further study. Regarding this
issue, we observe \cite{Glazek2011} the following:
(1) The AdS 5th dimension appears to correspond to
the quark size $s$ in RGPEP; (2) $G$-induced
oscillator potentials appear to correspond to the
soft-wall (SW) models \cite{SW}; (3) the resulting
identification $\kappa^2_{SW} = 2m \omega_M = (2
\pi/3) \, \varphi_G$ suggests that the soft wall
in the SW models results from the gluon condensation 
in hadrons; (4) AdS/QCD SW phenomenology result 
$ {\kappa_M /\kappa_B } \sim 1.15 \pm 0.5 $ matches 
our result for the same quantity, $ \left({8/5}\right)^{1/4} 
\sim 1.125 $. These results support the idea that the 
condensates associated with the vacuum state in the 
IF of dynamics are actually associated only with the 
hadronic interior \cite{BrodskyShrock,BrodskyRobertsShrockTandy}
in the FF of Hamiltonian dynamics. 

The last comment we wish to make here is that the
hadron content-induced harmonic oscillator potential 
between quarks at large RGPEP size $s$ implies LF
eigenvalues $M^2 \sim r^2$ for states where $r$ is 
large. This quadratic behavior of $M^2$ implies 
$M \sim r $, which is in agreement with the Regge 
phenomenology and string picture of hadrons.

\begin{acknowledgements}
I would like to thank Simon Dalley and the Organizing 
Committee for the invitation and opportunity to 
participate in the excellent meeting. 
\end{acknowledgements}


\begin{thebibliography}{3}

\bibitem[Dirac P (1949)]{Dirac}
Dirac P (1949) Forms of Relativistic Dynamics.
Rev. Mod. Phys. 21: 392--399 

\bibitem[ Dirac P (1965)]{DiracDeadWood}
Dirac P (1965) Quantum Electrodynamics without Dead Wood.
Phys. Rev. 139: B684--B690

\bibitem[Wilson K (1965)]{Wilson1965}
Wilson K (1965) Model Hamiltonians for Quantum Field Theory. 
Phys. Rev. 140: B445--B457

\bibitem[Wilson K (1970)]{Wilson1970}
Wilson K (1970) Model of Coupling-Constant Renormalization.
Phys. Rev. D 2: 1438--1472 

\bibitem[Lepage G and Brodsky S (1980)]{LepageBrodsky}
Lepage G and Brodsky S (1980) 
Exclusive Processes in Perturbative Quantum Chromodynamics.
Phys. Rev. D 22: 2157--2198

\bibitem[Srivastava P and Brodsky S (2001)]{SrivastavaBrodsky}
Srivastava P and Brodsky S (2001) 
Light front quantized QCD in light cone gauge.
Phys. Rev. D 64: 045006--15 

\bibitem[ Gross D and Wilczek F (1973)]{af1}
Gross D and Wilczek F (1973) 
Ultraviolet Behavior of Non-Abelian Gauge Theories.
Phys. Rev. Lett. 30: 1343--1346  

\bibitem[Politzer H (1973)]{af2}
Politzer H (1973) 
Reliable Perturbative Results for Strong
Interactions?
Phys. Rev. Lett. 30: 1346--1349 

\bibitem[G{\l}azek S (2001)]{gluons}
G{\l}azek S (2001)
Dynamics of effective gluons.
Phys. Rev. D 63: 116006--18 

\bibitem[Shifman M et al. (1979a)]{SVZ1}
Shifman M et al. (1979a)
QCD and Resonance Physics. Sum Rules.
Nucl. Phys. B 147: 385--447  

\bibitem[Shifman M et al. (1979b)]{SVZ2}
Shifman M et al. (1979b)
QCD and Resonance Physics: Applications.
Nucl. Phys. B 147: 448--518 

\bibitem[G{\l}azek S and Wilson K (1993)]{SRG1}
G{\l}azek S and Wilson K (1993)
Renormalization of Hamiltonians.
Phys. Rev. D 48, 5863--5872

\bibitem[G{\l}azek S and Wilson K (1993)]{SRG2}
G{\l}azek S and Wilson K (1994)
Perturbative renormalization group for Hamiltonians.
Phys. Rev. D 49: 4214--4218

\bibitem[Wilson K et al. (1994)]{Wilson1994}
Wilson K et al. (1994) Non-perturbative QCD: A
weak-coupling treatment on the light front.
Phys. Rev. D 49: 6720--6766 

\bibitem[G{\l}azek S (2011)]{Glazek2011}
G{\l}azek S (2011) Reinterpretation of gluon condensate
in dynamics of hadronic constituents. 
Acta Phys. Polon. B 42: 1933--2010

\bibitem[G{\l}azek S (2010)]{Glazek2010}
G{\l}azek S (2010) Non-local interactions in renormalized Hamiltonians. 
Acta Phys. Polon. B 41: 2669--2683

\bibitem[Wegner F (1994)]{Wegner}
Wegner F (1994) Flow-equations for Hamiltonians.
Ann. Phys. (Leipzig) 3, 77--91 

\bibitem[Kehrein S (2006)]{Kehrein}
Kehrein S (2006) 
The Flow Equation Approach to Many-Particle Systems.
Springer Verlag, Berlin

\bibitem[Glazek S and Schaden M (1987)]{GlazekSchaden}
G{\l}azek S and Schaden M (1987)
Gluon condensate induced confinement in mesons and baryons.
Phys. Lett. B 198: 42-44 

\bibitem[Isgur N and Karl G (1979)]{CQM1}
Isgur N and Karl G (1979)
Ground-state baryons in a quark model with hyperfine interactions.
Phys. Rev. D 20: 1191–1194 

\bibitem[Murthy M at al. (1984)]{CQM2}
Murthy M at al. (1984)
Rotational bands in the baryon spectrum. II.
Phys. Rev. D 30: 152–162 

\bibitem[Brauer K at al. (1985)]{CQM3}
Brauer K at al. (1985)
On one pion exchange potential with quark exchange in the resonating group method. 
Z. Phys. A 320: 609-612 

\bibitem[Brodsky S and Teramond G (2008)]{FAdS1}
Brodsky S and Teramond G (2008a)
Light-front dynamics and AdS/QCD correspondence: 
The pion form factor in the space- and time-like
regions.
Phys. Rev. D 77: 056007-20 

\bibitem[Brodsky S and Teramond G (2008b)]{FAdS2}
Brodsky S and Teramond G (2008b)
Light-front dynamics and AdS/QCD correspondence: 
Gravitational form factors of composite hadrons.
Phys. Rev. D 78: 025032-16

\bibitem[Brodsky S and Teramond G (2009)]{FAdS3}
Brodsky S and Teramond G (2009) 
Light-Front Holography: A First Approximation to QCD.
Phys. Rev. Lett. 102: 081601-4

\bibitem[Karch A et al. (2006)]{SW}
Karch A et al. (2006)
Linear confinement and AdS/QCD.
Phys. Rev. D 74: 015005-7 

\bibitem[Brodsky S and Shrock R (2011)]{BrodskyShrock}
Brodsky S and Shrock R (2011)
Condensates in quantum chromodynamics and the cosmological constant.
Proc. Natl. Acad. Sci. USA 108: 45-50 

\bibitem[Brodsky S at al. (2010)]{BrodskyRobertsShrockTandy}
Brodsky S at al (2010)
New perspectives on the quark condensate.
Phys. Rev. C 82: 022201(R)-5 

\end{thebibliography}
\end{document}